\documentclass[%
reprint,
superscriptaddress,
amsmath,
amssymb,
aps,
prapplied,
floatfix,
citeautoscript,
longbibliography
]{revtex4-2}

\usepackage{preamble}

\graphicspath{{./img/}}

\begin{document}

\title{Overcoming Noise Limitations in QKD with Quantum Privacy Amplification}

\author{Philipp Sohr}
\email{philipp.sohr@qtlabs.at}
\affiliation{Institute for Quantum Optics and Quantum Information (IQOQI), Austrian Academy of Sciences, Boltzmanngasse 3, 1090 Vienna, Austria}
\affiliation{Atominstitut, Technische Universit\"at Wien, 1020 Vienna, Austria}
\affiliation{Vienna Center for Quantum Science and Technology (VCQ), Faculty of Physics, University of Vienna, Boltzmanngasse 5, 1090 Vienna, Austria}
\affiliation{Quantum Technology Laboratories GmbH, Clemens-Holzmeister-stra\ss{}e 6/6, 1100 Vienna, Austria}

\author{Sebastian Ecker}
\affiliation{Institute for Quantum Optics and Quantum Information (IQOQI), Austrian Academy of Sciences, Boltzmanngasse 3, 1090 Vienna, Austria}
\affiliation{Vienna Center for Quantum Science and Technology (VCQ), Faculty of Physics, University of Vienna, Boltzmanngasse 5, 1090 Vienna, Austria}
\affiliation{Quantum Technology Laboratories GmbH, Clemens-Holzmeister-stra\ss{}e 6/6, 1100 Vienna, Austria}

\author{Lukas Bulla}
\affiliation{Institute for Quantum Optics and Quantum Information (IQOQI), Austrian Academy of Sciences, Boltzmanngasse 3, 1090 Vienna, Austria}
\affiliation{Vienna Center for Quantum Science and Technology (VCQ), Faculty of Physics, University of Vienna, Boltzmanngasse 5, 1090 Vienna, Austria}
\affiliation{Quantum Technology Laboratories GmbH, Clemens-Holzmeister-stra\ss{}e 6/6, 1100 Vienna, Austria}

\author{Martin Bohmann}
\affiliation{Institute for Quantum Optics and Quantum Information (IQOQI), Austrian Academy of Sciences, Boltzmanngasse 3, 1090 Vienna, Austria}
\affiliation{Vienna Center for Quantum Science and Technology (VCQ), Faculty of Physics, University of Vienna, Boltzmanngasse 5, 1090 Vienna, Austria}
\affiliation{Quantum Technology Laboratories GmbH, Clemens-Holzmeister-stra\ss{}e 6/6, 1100 Vienna, Austria}

\author{Rupert Ursin}
\affiliation{Institute for Quantum Optics and Quantum Information (IQOQI), Austrian Academy of Sciences, Boltzmanngasse 3, 1090 Vienna, Austria}
\affiliation{Vienna Center for Quantum Science and Technology (VCQ), Faculty of Physics, University of Vienna, Boltzmanngasse 5, 1090 Vienna, Austria}
\affiliation{Quantum Technology Laboratories GmbH, Clemens-Holzmeister-stra\ss{}e 6/6, 1100 Vienna, Austria}

\begin{abstract}
    High-quality, distributed quantum entanglement is the distinctive resource for quantum communication and forms the foundation for the unequalled level of security that can be assured in quantum key distribution.
    While the entanglement provider does not need to be trusted, the secure key rate drops to zero if the entanglement used is too noisy.
    In this paper, we show experimentally that QPA is able to increase the secure key rate achievable with QKD by improving the quality of distributed entanglement, thus increasing the quantum advantage in QKD.
    Beyond that, we show that QPA enables key generation at noise levels that previously prevented key generation.
    These remarkable results were only made possible by the efficient implementation exploiting hyperentanglement in the polarisation and energy-time degrees of freedom.
    We provide a detailed characterisation of the gain in secure key rate achieved in our proof-of-principle experiment at different noise levels.
    The results are paramount for the implementation of a global quantum network linking quantum processors and ensuring future-proof data security.
\end{abstract}

\maketitle

\section{Introduction}
\begin{figure*}[t]
  \centering
  \includegraphics[width=0.8\linewidth]{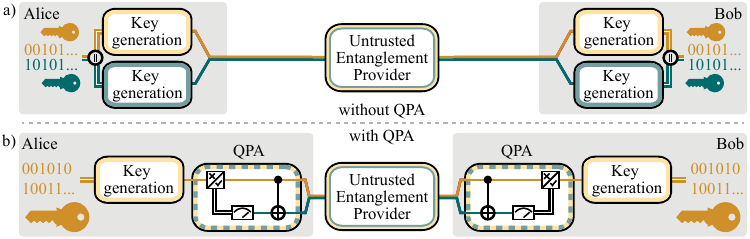}
  \caption{
    Schematic of entanglement distribution for QKD with and without quantum privacy amplification~(QPA).
    Alice and Bob are provided with hyperentangled states by an untrusted, potentially malicious party that controls both the entanglement source and the quantum channel.
    The two degrees of freedom~(DOF) are indicated by the yellow and turquoise colours, respectively.
    In scenario (a), a separate key is generated independently from each degree of freedom first.
    Then, the two keys are concatenated, which increases the total key rate.
    However, the distributed entanglement could be noisy and mixed, either due to an eavesdropping attempt by the entanglement provider or due to interaction with the environment during distribution.
    This reduces the secure key rate and as soon as the noise threshold of the employed QKD protocol is reached, the secure key rate drops to zero, i.e. no secure key can be generated.
    With QPA as shown in scenario (b), the noise threshold can be increased and a key can be generated in noise ranges that are not accessible in scenario (a).
    Alice and Bob each perform a controlled NOT gate between the DOF of their received photon.
    They project the target DOF~(turquoise) onto the computational basis.
    If their measurement results agree, they keep the photon pair.
    Otherwise, they discard the photon pair.
    The entanglement in the remaining DOF is used as a resource to generate a key.
    With the quantum resources enhanced by QPA, it is possible to raise the noise threshold and achieve key rates in the one remaining DOF that are higher than the concatenated key rate without QPA.
    }
  \label{fig:scheme}
\end{figure*}
Quantum key distribution~(QKD) allows two communication parties, Alice and Bob, to generate a cryptographic key at a distance, even in the presence of a technologically unbounded adversary~\cite{Scarani2009, Renner2022}.
The distinctive feature of QKD is the security based on physical principles~\cite{Renner2005, Portmann2021}.
Since the security of QKD can be proven without making assumptions on the adversary whatsoever, the security is not threatened by technological progress of the adversary.

Quantum entanglement is one of the outstanding physical phenomena used as a resource for quantum information processing tasks.
By measuring entangled photon pairs at distant locations, identical randomness is generated at these locations, which is exclusively shared by two parties~\cite{Horodecki2009a, Zeilinger1999}.

In fact, these characteristic properties of entanglement are precisely those that are essential for the distribution of a symmetric cryptographic key~\cite{Curty2004}. 
Entanglement-based QKD protocols such as E91~\cite{Ekert1991} and BBM92~\cite{Bennett1992a} make use of these properties.
As long as Alice and Bob share pure entangled states, the monogamy of entanglement guarantees privacy of the resulting randomness~\cite{Koashi2004, Barrett2006a}.
In the absence of errors caused by environmental noise, adversarial interaction or device imperfections, a perfectly secure key would be shared between Alice and Bob already after sifting of the measured data.
However, entanglement cannot be generated remotely by local operations and classical communication~(LOCC), but it has to be generated in one place to be then distributed to the communication parties.
Inevitable interaction with the environment during distribution as well as eavesdropping attempts reduce the purity, leaving Alice and Bob with noisy, mixed entangled states.
As a result, Alice and Bob share nonperfect correlations after measurement and sifting.
By analysing the measurement data, it is possible to monitor the correlations resulting from entanglement and thereby detect any eavesdropping attempts~\cite{Bennett1992a}.
Due to this unique feature of identifying eavesdropping attempts, Alice and Bob do not need to trust the entanglement provider.
It is therefore conceivable that not only the quantum channel but also the source of the entangled photons is under the control of a malicious third party.

Whenever Alice and Bob discover an eavesdropping attempt, they take appropriate measures to assure the security of distributed keys.
In classical postprocessing, the security of the distributed key is increased at the cost of the length of the key.
If the entanglement is too noisy, whether due to cheating by the entanglement provider or interaction with the environment during transmission, the secure key length is reduced to zero~\cite{Bennett1995, Renner2005c}.
But even if this noise threshold is exceeded, it is still possible to generate a secure key with QKD by improving the quantum resources before the classical postprocessing.

One approach to increase the noise resistance~\cite{Ecker2019} is high-dimensional~(high-dim) QKD, that uses high-dim entanglement as an advanced quantum resource~\cite{Erhard2020, Cozzolino2019b, Hu2021a}.
With increasing dimension, the background noise is more and more diluted, leading to a higher signal-to-noise ratio and an increased noise threshold~\cite{Sheridan2010a, Ferenczi2012a}.
While high-dim entanglement is readily produced in experiment, the measurements required for high-dim QKD are involved~\cite{Bulla2023a, Bulla2023b, Bergmayr2023}.
Another approach is entanglement distillation~\cite{Bennett1996a}, which can improve the shared entanglement by LOCC at the cost of the number of qubit pairs.
In the context of QKD, as shown in Fig.~\ref{fig:scheme}, entanglement distillation is known as Quantum Privacy Amplification~(QPA)~\cite{Deutsch1996}.
In the original QPA proposal, two photon pairs that are sufficiently~\cite{Horodecki1996} but not maximally entangled were used to output a single photon pair closer to a maximally entangled state.
However, implementations of a two-copy scheme are faced with two major problems:
(i) the impossibility of implementing a deterministic controlled NOT~(CNOT) operation between two independent photons with passive linear optics~\cite{OBrien2003, Pittman2003, Gasparoni2004, Zhao2005} and finite resources and 
(ii) the low probability of simultaneous transmission of two photon pairs.
Recently, it has been shown that single-copy entanglement distillation can leverage the entanglement of different degrees of freedom~(DOF) of a single photon pair to overcome these challenges~\cite{Ecker2021a, Hu2021b}.
Similarly to high-dim entanglement, hyperentanglement is readily produced in spontaneous parametric down-conversion~(SPDC)~\cite{Barreiro2005, Kwiat1997} and with other photon-pair generation techniques~\cite{Prilmueller2018, Reimer2019}.
However, while the exploitation of high-dim entanglement for QKD requires sophisticated measurements, QKD in conjunction with QPA, in contrast, is performed with common qubit measurements.
Notably, the deterministic realisation of the CNOT gate between two DOF of the same photon, as shown in Fig.~\ref{fig:scheme}, is possible with linear optics~\cite{Fiorentino2004, Barreiro2008}. %
The polarisation~(pol) and energy-time~(e-t) DOF employed by the work presented in Ref.~\cite{Ecker2021a} are known for their robustness as quantum information carriers and have been successfully distributed over free space~\cite{Yin2017, Steinlechner2017, Jin2019, Bulla2023a} and long-haul optical fibre links~\cite{Wengerowsky2019, Marcikic2004, Neumann2022b}, making them an ideal choice for future applications.
It has been demonstrated, that single-copy entanglement distillation can improve the Bell-state fidelity for a wide range of noisy input states ~\cite{Ecker2021a}.
For three single noise levels and using the less practical spatial encoding, a potential advantage of a single-copy distillation experiment has been shown~\cite{Hu2021b}.
However, it remains to be shown that the effects of the increased entanglement quality outweigh the reduced number of entangled qubit pairs, resulting in an actual advantage of QPA for QKD.

Here, we report on the experimental demonstration of QPA with hyperentanglement in the field-tested polarisation and energy-time degrees of freedom.
We compare the sum of the key rates extracted from both DOF before QPA with the key rate extracted from the polarisation DOF after QPA, see Fig.~\ref{fig:scheme}.
We show that the implementation of QPA not only increases the overall key rate, but also enables QKD with noisy entanglement in cases where this would no longer be possible with classical postprocessing alone.
To fully explore this process, we intentionally generate noisy entanglement in a finely controlled manner by mixing two Bell states in the polarisation DOF and three Bell states in the energy-time DOF.
We systematically assemble a range of noise contributions into a map to precisely show the performance of our setup under various conditions.

\section{Experiment}
\begin{figure*}[t]
  \centering
  \includegraphics[width=0.8\linewidth]{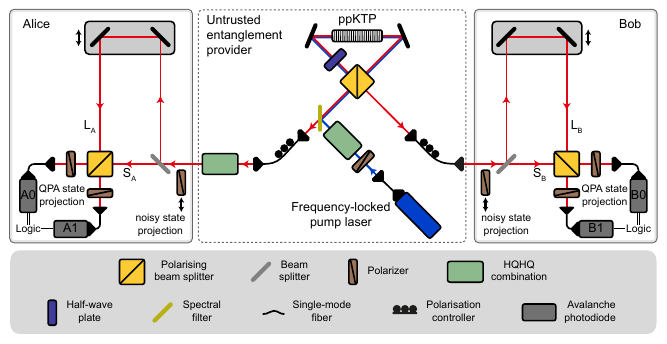}
  \caption{
    Experimental setup scheme of Quantum Privacy Amplification.
    Pairs of single photons hyperentangled in the polarisation and the energy-time degree of freedom~(DOF) are created in a spontaneous parametric down conversion~(SPDC) process by bidirectional pumping of a periodically poled potassium titanyl phosphate~(ppKTP) crystal in a Sagnac-type source with a \qty{405}{nm} narrow-bandwidth continuous-wave laser.
    The single photons are coupled to a single-mode fibre and distributed to Alice and Bob.
    Combinations of two quarter- and two half-wave plates~(HQHQ) are used to tune the state generated at the source as well as to create mixtures of Bell states.
    To perform the QPA step, both Alice and Bob are equipped with an imbalanced Mach-Zehnder interferometer each with a polarising beam splitter~(PBS) at the output, constituting a Franson-type interferometer.
    Each PBS acts as deterministic controlled NOT gate between the polarisation DOF and the energy-time (path) DOF of one and the same photon.
    The translation stages in the long arm are used both to adjust the phase of the energy-time Bell state, and to counteract length changes of the interferometer arms due to mechanical influences and temperature drifts in the laboratory.
    The single photons are detected by fibre-coupled avalanche photodiodes.
    The measurements for key estimation are performed with polarisation filters after and optionally also before the interferometer.
    In the energy-time DOF, this projective measurement corresponds to the identification of the detector that detected a photon.
    For the sake of clarity, auxiliary systems such as the stabilisation of the Franson interferometer are omitted in this figure. A more detailed scheme can be found in Figure 2 of Ref. \cite{Ecker2021a}.
    }
  \label{fig:setup}
\end{figure*}
We generate hyperentangled photon pairs in an SPDC process.
A Franson-type interferometer, as shown in Fig.~\ref{fig:setup}, projects the energy-time DOF to a 2-dimensional path DOF, such that a polarising beam splitter~(PBS) can act as CNOT gate flipping the path of the photon depending on the polarisation.
Postselection on the path modes after the Franson-type interferometer completes QPA.
The performance of the setup is quantified by local projective measurements, both before and after the interferometer.

In our table-top experiment, we had full control over the source of entangled photons as well as over the noisy channel.
The high-quality polarisation entanglement is generated by carefully tuning the polarising Sagnac interferometer~\cite{Kim2006, Fedrizzi2007} to maximise the overlap of the biphoton modes resulting from the bidirectional pumping of the ppKTP crystal.
The energy-time entanglement~\cite{Martin2017} arises naturally by pumping the crystal with a narrow-band continuous-wave laser.
While it is known that the generation time of the single photons of a pair must be the same due to the conservation of energy, the absolute generation time of a pair cannot be predicted. 
All generation times within the coherence time of the laser are coherently superimposed.
By the Franson-type interferometer~\cite{Franson1989}, this high-dimensional entanglement is projected on a 2-dimensional path DOF, namely the short~(S) and the long~(L) paths of the interferometer, each with a delay $t_\mathrm{S_\mathrm{A/B}}$ and $t_\mathrm{L_\mathrm{A/B}}$, respectively.
The phase of the energy-time Bell state is tuned by carefully adjusting the length imbalance of one of the Mach-Zehnder interferometers on the order of the wavelength, see Figure~\ref{fig:setup}.
As a result, we obtain a hyperentangled state $\Phi^+$, that is close to a $\phi^+$ Bell state in both DOF
\begin{equation}
    \left|\Phi^+\right\rangle = \frac{1}{2}\left[
    \left(\left|H, H\right\rangle + \left|V, V\right\rangle\right)
    \otimes
    \left(\left|t_\mathrm{S}, t_\mathrm{S}\right\rangle + \left|t_\mathrm{L}, t_\mathrm{L}\right\rangle\right)
    \right]\,.
\end{equation}

The mixed state in the polarisation DOF is generated with a combination of wave plates acting on one photon of a pair.
The wave plates are adjusted in a way that their combination causes a tunable rotation about the $y$ axis on the Bloch sphere $\left(R_\mathrm{y}(\theta) \otimes \mathbb{1}\right)$.
By time averaging over the settings for $\pm\theta$, the Bell state $\psi^-$ is mixed to the originally prepared $\phi^+$ state.
In the energy-time DOF, the mixed state is generated by increasing the coincidence window for accepting coincidences in the postprocessing.
By including the noninterfering background of the Franson interference, an equal admixture of the Bell states $\psi^+$ and $\psi^-$ to the originally prepared $\phi^+$ state is achieved (for details see Appendix of Ref. \cite{Ecker2021a}).

The heart of the QPA setup (Fig.~\ref{fig:setup}) are the two PBSs acting as bilateral CNOT between the DOF of one and the same photon.
The coincidence events are accepted if either detectors A0 and B0 or detectors A1 and B1 registered a photon, otherwise, they are discarded.

The ratio between the accepted coincidences $C_\mathrm{pass}$ and the total coincidences $C_\mathrm{tot}$ is defined as the yield of the QPA process $y = C_\mathrm{pass} / C_\mathrm{tot}\,.$
In contrast to the originally proposed two-copy QPA~\cite{Deutsch1996}, which is limited to a maximal yield of $0.5$, since one photon pair was always consumed independent of the quality of the input states, the yield of the single-copy QPA presented here approaches $1$.
Even though the fidelity with respect to a Bell state can be increased with such a setup, with a yield up to one~\cite{Ecker2021a}, it is not obvious, that this manifests a benefit for QKD.

To estimate a lower bound on the key rate, we measure in the two linear mutually unbiased bases, i.e. the computational basis $\left\lbrace\left|H\right\rangle, \left|V\right\rangle\right\rbrace$ in the polarisation DOF and $\left\lbrace\left|t_\mathrm{S}\right\rangle, \left|t_\mathrm{L}\right\rangle\right\rbrace$ in the energy-time DOF as well as the superposition bases $\left\lbrace\left(\left|H\right\rangle \pm \left|V\right\rangle\right)/\sqrt{2}\right\rbrace$ and $\left\lbrace\left(\left|t_\mathrm{S}\right\rangle\pm \left|t_\mathrm{L}\right\rangle\right)/\sqrt{2}\right\rbrace$, respectively.
The noisy entanglement, introduced by the admixture of Bell states, is recognised as erroneous detection events in the experiment $N_\mathrm{err}$.
The noise is quantified by the quantum bit error rate~(QBER), which is defined as the quotient of the number of erroneous detection events and the total number of events, $e = N_\mathrm{err}/N$.
With this at hand, we compute a lower bound of the secure key rate $k$ in the asymptotic limit of infinite key length with the Devetak-Winter formula~\cite{Devetak2005}
\begin{equation}
    \label{eq:secureKeyRate}
    k\left(e_\mathrm{z}, e_\mathrm{x}\right) \geq \mathrm{max}\left(0, 1 - h_2(e_\mathrm{z}) - h_2(e_\mathrm{x})\right)\,,
\end{equation}
where $h_2\left(x\right) = - x \mathrm{log}_2 \left( x \right) - \left( 1 - x \right) \mathrm{log}_2 \left( 1 - x \right)$ is the binary Shannon entropy.
Before the QPA step, both DOF of the hyperentangled state can in principle be used to generate a secure key as depicted in Figure~\ref{fig:scheme}.
The secure key rate before the QPA $k_\mathrm{noisy}$ is the sum of the secure key rate in each DOF,
\begin{align}  
    \label{eq:secKey-noisy}
        k_\mathrm{noisy} 
        = k\left(e_\mathrm{z}^\mathrm{pol}, e_\mathrm{x}^\mathrm{pol}\right) + \frac{k\left(e_\mathrm{z}^\mathrm{e-t}, e_\mathrm{x}^\mathrm{e-t}\right)}{2}\,.
\end{align}
After QPA, the QBER in the polarisation DOF $\check{e}_\mathrm{x/z}^\mathrm{pol}$ has changed and the yield $y$ has to be taken into account, so that the key rate after QPA $k_\mathrm{QPA}$ computes as
\begin{equation}
    \label{eq:secKey-QPA}
    k_\mathrm{QPA} = y * \frac{k\left(\check{e}_\mathrm{z}^\mathrm{pol}, \check{e}_\mathrm{x}^\mathrm{pol}\right)}{2}\,.
\end{equation}
The factor $1/2$ in Equations \ref{eq:secKey-noisy} and \ref{eq:secKey-QPA} reflects the $50\%$ loss due to the Franson interferometer.
As a measure for the performance of the protocol, we define the gain $g$ in secure key rate as the difference between the sum of the key rate after QPA and the two key rates using noisy entanglement
\begin{equation}
    \label{eq:gain}
    g = k_\mathrm{QPA} - k_\mathrm{noisy}\,.
\end{equation}

\begin{figure}[H]
  \centering
  \includegraphics[width=\columnwidth]{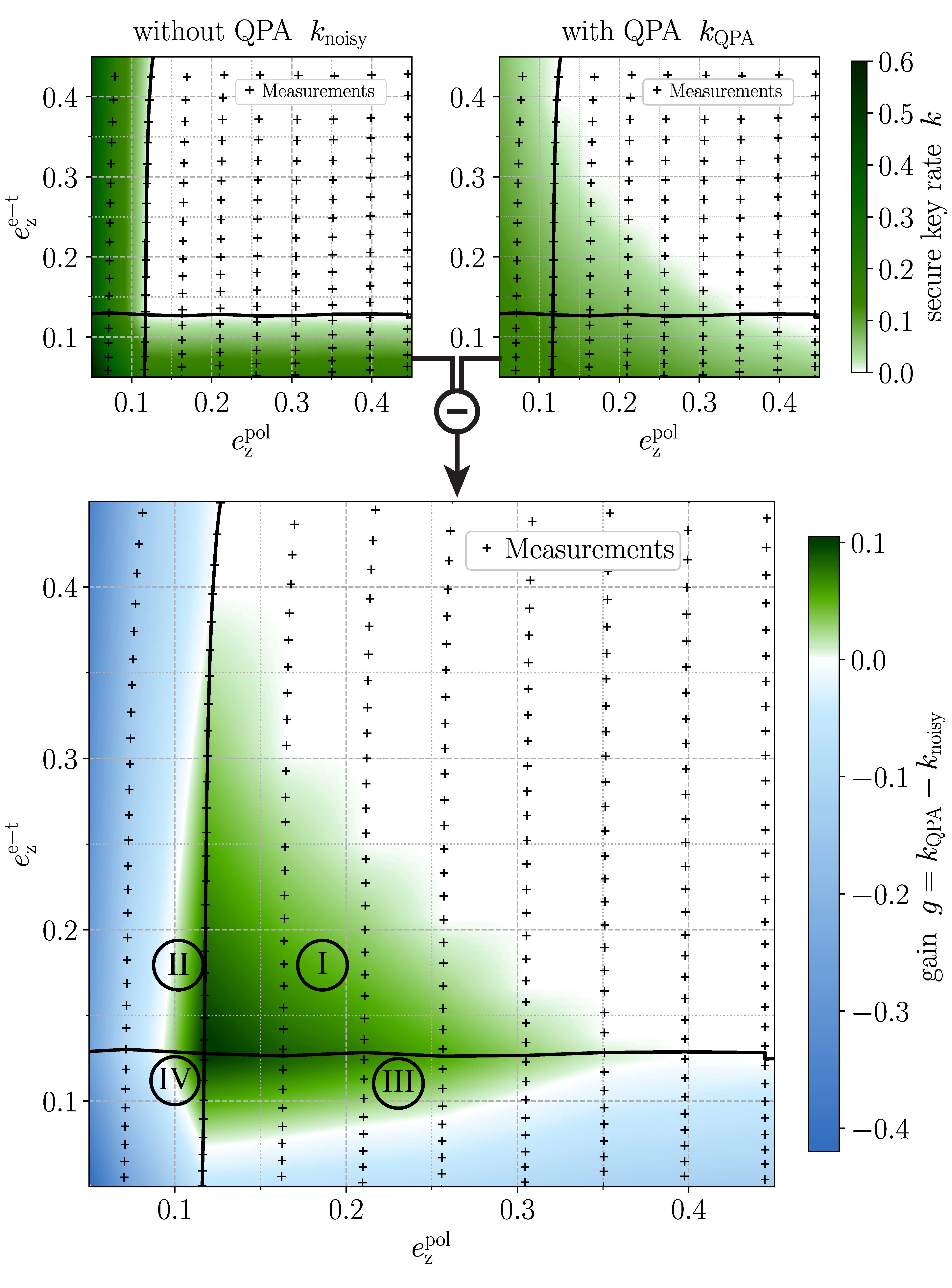}
  \caption{
  Advantage in the secure key rate by QPA mapped for various noise levels in each DOF separately.
  The measurement points indicated by the markers (not all are shown for readability; the error bars are smaller than the markers) span a grid of the qber in the $z$ basis in both DOF.
  In the two top plots, the colour, that results from a linear interpolation between the measurement points, indicates the secure key rate before and after the QPA step.
  Both heatmaps share the same colorbar.
  The thick black lines indicate the qber threshold, which is about $11\%$ in the polarisation DOF and slightly higher in the energy-time DOF due to the noise composition.
  Comparing the two heatmaps, it is already clearly visible that the area of the positive key rate with QPA extends over the limits of the key rate without QPA.
  Even clearer becomes the advantage of QPA in the bottom heatmap displaying the QPA gain in the secure key rate.
  The region of positive gain reaches down to $9.1\%$ qber in the polarisation DOF and $7.2\%$ qber in the energy-time DOF up to $40\%$ qber in either DOF.
  We reach a maximum gain of 0.105.
  Not only is it possible to generate a key where it was not possible before (I), but also can the key rate be improved in regions where key generation was possible in one DOF (II), (III) and even in both DOF (IV) before.
  }
  \label{fig:sec-key}
\end{figure}

\section{Results}

We obtained a true advantage by increasing the noise threshold of QKD with our QPA experiment, indicated by the positive gain shown in Figure~\ref{fig:sec-key}.
The region of positive gain can be divided into three parts.
Firstly, the regions (II) and (III), where the noise threshold was exceeded in one DOF, but secure key generation was still possible in the other DOF before the QPA step.
In these regions the application of QPA increased the secure key rate.
Region (III) is larger than region (II) due to the $50\%$ loss introduced by the implementation of the Franson interferometer to utilize the energy-time entanglement.
Secondly, in a small domain (IV) the noise was close to the threshold in both DOF before the QPA.
With QPA, we increased the secure key rate in this region even though both DOF contributed to the total pre-QPA secure key rate.
Thirdly, in the largest part of the region of positive gain (I), we generated a secure key rate facing a noise level that prevented key generation before.
This shows that QPA can increase the noise threshold of QKD, enabling a positive secure key rate where this was not possible before.
Negative gains are obtained if high-quality entanglement is provided at least in one of the two DOF. 
While the maximal gains will always be found along the noise thresholds, the amplitude and spread of negative gain can be reduced by improving the implementation of QPA.

\section{Discussion and Conclusions}
With this work, we demonstrate experimentally that QPA boosts the quantum advantage of QKD.
Provided with noisy quantum resources, Alice and Bob usually process their classical records of the quantum measurements to obtain identical, secret keys.
However, once the noise threshold of the employed protocol is reached, the secure key length reduces to zero, rendering QKD impossible with the provided resources.
We show that QPA can improve the quality of the quantum resources, such that the post-QPA key rate exceeds the total pre-QPA key rate close to and beyond the noise threshold of QKD.
This pioneering demonstration was only made possible by the increased efficiency of the single-copy scheme harnessing hyperentanglement in the polarisation and the energy-time DOF.
Not only does the hyperentanglement enable an implementation of a deterministic CNOT gate with linear optics without ancillary photons, but also is the efficiency increased since each QPA step requires the distribution of only one photon pair.
As a consequence of the latter, our QPA method has quadratic advantage in both the creation and the transmission probability of the photon pairs compared to two-copy schemes.
This advantage is particularly relevant in high-loss regimes such as long-distance links. 
The constant $\qty{2}{dB}$ loss introduced by the Franson interferometer can be avoided by active switching~\cite{Vedovato2018}.
The dense grid of measurement points for various noise levels in both DOF subspaces independently does allow a thorough assessment of the implemented QPA scheme's performance.
The resulting map of the advantage of QPA is not only of great importance for implementation purposes, but also for further focused research.
The most remarkable finding of this work is that QPA can increase the noise threshold, enabling QKD in hitherto inaccessible noise regimes.
While the noise threshold can also be raised with high-dim QKD, QKD supported by QPA works with qubit measurements that are easier to implement and the QPA module can even be retrofitted in deployed QKD systems.
Combining the noise advantages of high-dim QKD with our QPA method promises robustness to various types of noise.
We have demonstrated the capabilities of QPA to counteract a noise factorising in the DOF subspaces.
Isotropic noise in the joint state space of different DOF such as noise caused by background light or detector dark counts can be diluted by embedding the polarisation state in a high-dimensional state space, such as the energy-time state space~\cite{Ecker2019}.
The results of this work are of great importance for implementations of QKD, a technology that is currently emerging from fundamental research toward commercial application.
As such, our results will contribute significantly to the advent of a robust global QKD network providing secure communication and forming the basis of the quantum internet.
In the future, implementations of QPA could be extended to further DOF, given that an efficient CNOT gate can be performed between them.
Using entanglement in additional DOF could enable more QPA steps with a single photon pair, further increasing the advantage of QPA.
In particular, this would allow to counteract arbitrary subspace noise~\cite{Bennett1996a}.
Finally, it remains open to demonstrate the advantage of QPA in an in-field test with real noise as well as the combination with high-dim QKD schemes.

\section*{Data availability}

The experimental data and the analysis will be provided upon reasonable request.

\section*{Acknowledgements}
We acknowledge funding from the Austrian Science Fund (FWF) through the START project Y879-N27 and from the European  Unions  Horizon  2020  programme grant agreement No. 857156 (OpenQKD).

P.S., S.E. and R.U. conceived the project; P.S., S.E. and L.B. designed and developed the experiment under the guidance of M.B. and R.U.; P.S. and S.E. evaluated the experimental data; P.S. wrote the first draft of the manuscript; all authors discussed the results, contributed to writing and reviewed the manuscript; M.B. and R.U. supervised the project.

\bibliography{references}
\end{document}